\documentclass[aps,prb,twocolumn,groupedaddress,showpacs,amssymb]{revtex4}
\usepackage{graphicx}
\usepackage{dcolumn}
\usepackage{bm}

\begin{document}

\preprint{}

\title{Heat Capacity in Magnetic and Electric Fields Near the Ferroelectric Transition in Tri-Glycine Sulfate }

\author{J.C. Lashley}
\affiliation{
   Los Alamos National Laboratory,
   Los Alamos, New Mexico 87545}

\author{C.P. Opeil}
\affiliation{
   Department of Physics,
Boston College, Chestnut Hill, MA 02467}

\author{T. R. Finlayson}
\affiliation{
   School of Physics, Monash University,
   Clayton, Victoria, Australia 3800}

\author{R.A. Fisher}
\affiliation{
   Lawrence Berkeley National Laboratory,
   Berkeley, California 94720}

\author{N. Hur}
\affiliation{Department of Physics, Inha University, Incheon
402-751, Korea}

\author{M.F. Hundley}
\affiliation{
   Los Alamos National Laboratory,
   Los Alamos, New Mexico 87545}

\author{B. Mihaila}
\affiliation{
   Los Alamos National Laboratory,
   Los Alamos, New Mexico 87545}

\author{J.L. Smith}
\affiliation{
   Los Alamos National Laboratory,
   Los Alamos, New Mexico 87545}

\begin{abstract}
Specific-heat measurements are reported near the Curie temperature
($T_C$~=~320~K) on tri-glycine sulfate. Measurements were made on
crystals whose surfaces were either non-grounded or short-circuited,
and were carried out in magnetic fields up to 9~T and electric
fields up to 220~V/cm. In non-grounded crystals we find that the
shape of the specific-heat anomaly near $T_C$ is thermally
broadened. However, the anomaly changes to the characteristic sharp
$\lambda$-shape expected for a continuous transition with the
application of  either a magnetic field or an electric field. In
crystals whose surfaces were short-circuited with gold, the
characteristic $\lambda$-shape appeared in the absence of an
external field. This effect enabled a determination of the critical
exponents above and below $T_C$, and may be understood on the basis
that the surface charge originating from the pyroelectric
coefficient, $dP/dT$, behaves as if shorted by external magnetic or
electric fields.
\end{abstract}

\maketitle

The discovery of a continuous-phase transition in tri-glycine
sulfate, $[(CH_2NH_2CO_2H)_3\cdot\,H_2SO_4]$ (TGS) showing the
dielectric analog of ferromagnetism, was made fifty-years ago at
Bell Laboratories~\cite{Matthias}. This transition was considered
ideal in that critical exponents, as determined from the electric
polarization, $P$, and the dielectric constant, $\epsilon$, are
equal both above and below $T_C$~\cite{Gonzalo66,Gonzalo}. Today
pyroelectric sensors based on TGS have technological importance for
the development of flat-panel displays~\cite{Rosenman} and
room-temperature detectors of far-infrared laser
radiation~\cite{Hadni1, Hadni2}. Their sensitivity originates from a
large pyroelectric coefficient~\cite{Hoshino},~$dP/dT$, low Curie
temperature, $T_C$~=~320~K, and low coercive electric
field~\cite{Matthias},~$E_c$=~220~V/cm.

\begin{figure}[!]
    \centering \includegraphics[width=0.9\columnwidth]{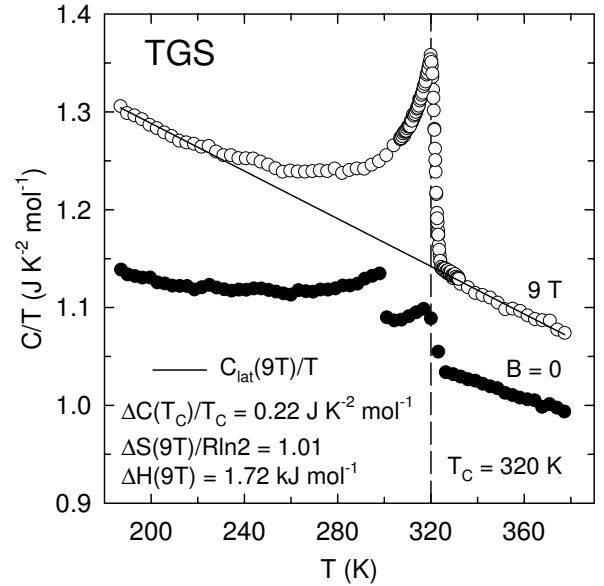}
    \caption{The temperature dependence of $C/T$ at 0~T (solid black circles) and 9~T (open circles)
    in the regime of $T_C$~=~320~K for a non-electrically grounded (free-standing) TGS single
    crystal. Data from measurements in fields as low as $3~T$, not shown, are congruent with those for $9~T$.
    }
    \label{Fig1}
\end{figure}

In an earlier review of the critical exponents for TGS,
Gonzalo~\cite{Gonzalo} pointed out that although they are
well-behaved above and below $T_C$ in $P$ and $\epsilon$
measurements, the paucity of data points in the vicinity of $T_C$
have prevented a determination using specific-heat data.
Consequently, many high-quality specific-heat measurements have been
made to characterize $T_C$ and to complete the set of critical
exponents for
TGS~\cite{Hoshino,Strukov,Taraskin,Shurmann,Joel,Ema,Boris76,Boris78,Boris80,Boris,Ramos,Ema2}.
Adiabatic measurements~\cite{Hoshino} made near $T_C$ on powdered
TGS samples were puzzling in the sense that for the first time, in
the field of ferroelectrics, the specific-heat anomaly for a
continuous transition was not of the $\lambda$ type. Instead the
transition was thermally broadened, giving an entropy change,
$\Delta S$~=~2~J~K$^{-1}$~mol$^{-1}$, that was much less than the
expected $R\ln2$~\cite{Mueller}. High-sensitivity ac measurements of
the Seidel type~\cite{George} made on short-circuited crystals,
showed a pronounced $\lambda$ anomaly at $T_C$~\cite{Joel}. Shortly
afterward, high-precision ac measurements reproduced the $\lambda$
anomaly, and the specific-heat data were shown to depend
logarithmically on $(T-T_C)$ in a range $T-T_C~\leq$~20~K~\cite{Ema}
above $T_C$. Seeking to clarify the experimental situation,
Strukov~\cite{Boris} performed high-accuracy adiabatic specific-heat
measurements and set limits on the relative measurement accuracy
needed for observing critical fluctuations near~$T_C$. In the
low-temperature limit, specific-heat measurements
~\cite{Lawless,Lawless2} demonstrated the surprising result that
surface excitations, (up to 250~microns thick) cause a $T^{3/2}$
term in the bulk heat capacity.

\begin{figure}[b]
    \centering \includegraphics[width=0.9\columnwidth]{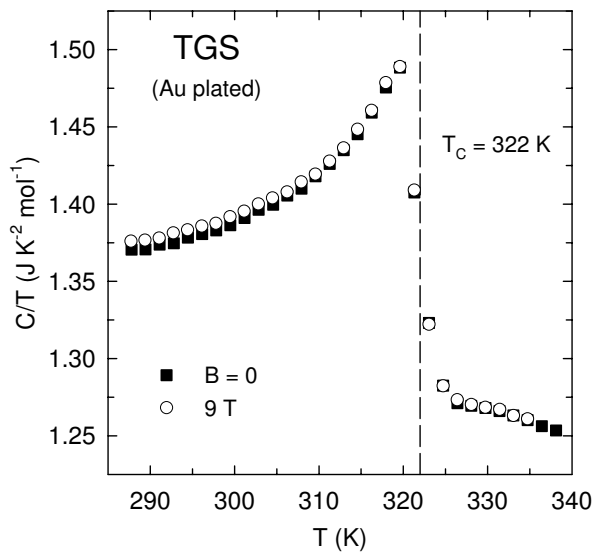}
    \caption{The temperature of $C/T$ near $T_C$ is shown for short-circuited (gold plated) TGS crystals
    at 0~T (solid squares) and 9~T (open circles). One sees that when the surface
    charge is shorted with a thin layer of gold, the characteristic
    $\lambda$-shape appears and shows no magnetic-field dependence.}
    \label{Fig2}
\end{figure}

It was therefore anticipated that because: (a)~surface excitations
are known to affect the bulk thermodynamic properties at
low-temperature giving rise to a $T^{3/2}$ term~\cite{Lawless2},
(b)~ non-grounded powders do not give the expected $\lambda$ anomaly
at $T_C$~\cite{Hoshino}, (c)~ac measurements show a pronounced
$\lambda$ anomaly with a logarithmic critical
fluctuations~\cite{Ema,Ema2}, and (d)~high-precision adiabatic
measurements show a smearing of the transition and do not have a
logarithmic singularity~\cite{Boris}, surface excitations generated
from the pyroelectric coefficient, $dP/dT$, were affecting the
nature of the specific-heat anomaly near~$T_C$.

In this Letter, we report measurements of the specific heat made on
crystals whose surfaces were either non-electrically grounded (free
standing) or short circuited with a thin layer of gold. The data for
non-electrically grounded crystals in zero magnetic or electric
fields show that the shape of the specific-heat anomaly near $T_C$
is thermally broadened, giving an entropy change smaller than the
expected~$R\ln2$, in agreement with previous results of Hoshino {\it
et al.}~\cite{Hoshino}. The anomaly changes into the characteristic
$\lambda$-shape expected for a continuous transition with the
application of  either magnetic or electric fields giving an
increased entropy change $\Delta
S$~=~5.8~J~K$^{-1}$~mol$^{-1}$~=~$R\ln2$. In crystals whose surfaces
were short-circuited with gold, the characteristic $\lambda$-shape
appeared at $T_C$ with no applied magnetic or electric fields.

The TGS crystals used for the experiments were grown from
supersaturated solutions of the components by evaporation. Crystal
quality was checked by examining the sharpness and reversibility of
the transition at $T_C$ as determined by variable-temperature
capacitance measurements~\cite{Daniels}. Specific-heat measurements
were made using a thermal-relaxation technique~\cite{Lashley}.
Thermal-relaxation calorimetry provides a means of determining a
sample's specific heat by measuring the thermal response of a
sample/calorimeter assembly to a change in heating conditions.
Crystal dimensions for the TGS used in the magnetic field
measurements were 2.85 mm x 3.78 mm x 1.11 mm with a mass of 17.89
mg. Crystal dimensions for the TGS used in electric field
measurements were 3.74 mm x 3.55 mm x 1.15 mm with a mass of 22.38
mg.

\begin{figure}[t]
    \centering \includegraphics[width=0.9\columnwidth]{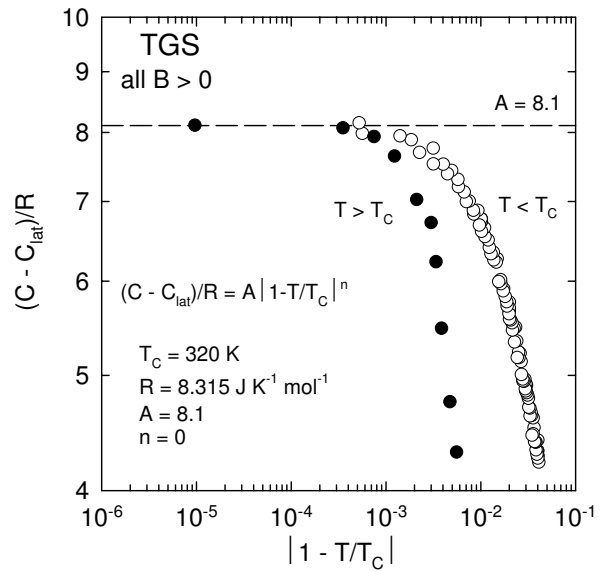}
    \caption{Log-log plot of the excess heat capacity versus reduced temperature showing the $\lambda$ anomaly above $T_C$ (solid
    circles) and below $T_C$ (open circles). Results for the critical
    exponents obtained on each side of $T_C$ are given in the plot
    legends. As an independent check, $RA/T_C$~=~0.21~J K$^{-2}$ mol$^{-1}$, which is equal to the value of the specific-heat discontinuity at $T_C$ given by $\Delta C (T_C)/T_C$, shown in Fig. 1}
    \label{Fig3}
\end{figure}

Figure~1 shows the temperature dependence of $C/T$ versus $T$ in the
vicinity of $T_C$, for $9~T$ on non-grounded (free-standing) TGS
crystals with the magnetic field applied parallel to the easy {\it
b}-axis. At zero magnetic field (solid circles), there is a broad
anomaly separated into two peaks near~$T_C$. These zero-field
results are strikingly similar to those reported nearly fifty years
ago by Hoshino {\it et al.}~\cite{Hoshino}. A previous theory
developed for Rochelle salt~\cite{Mueller}, showed that the free
energy for a continuous transition could be approximated by a
polynomial in the second and fourth power of the order parameter-the
electric polarization along the easy {\it b}-axis. Applying these
results to TGS, Hoshino {\it et al.}~\cite{Hoshino} found a linear
decrease of P$^2$ as a function of temperature approaching~$T_C$.
Using polarization and dielectric susceptibility data, Hoshino {\it
et al.}~\cite{Hoshino} were able to identify the numerical values of
the parameters in Mueller's model~\cite{Mueller}. However, the
experimental entropy change at $T_C$ was 33 percent lower than the
theoretical value. In those measurements the TGS was made into a
powder, and the sample was not electrically grounded. In agreement
with our zero-field results on a non-electrically grounded crystal,
their results had an entropy change of 2~J~K$^{-1}$mol$^{-1}$, and
did not have the characteristic $\lambda$ shape. After cooling the
non-grounded crystal back to room temperature and applying a 9~T
magnetic field at 300~K, the characteristic $\lambda$ shape
reappears (open circles), see Fig.~1. Integration of 9~T the $C/T$
versus $T$ curve gives the entropy change previously unaccounted for
in our zero magnetic field data in addition to the earlier data by
Hoshino {\it et al.}~\cite{Hoshino}, specifically $\Delta
S$~=~5.8~J~K$^{-1}$~mol$^{-1}$~=~$R\ln2$, as indicated in the legend
of Fig.~1.

\begin{figure}[t]
    \centering \includegraphics[width=0.9\columnwidth]{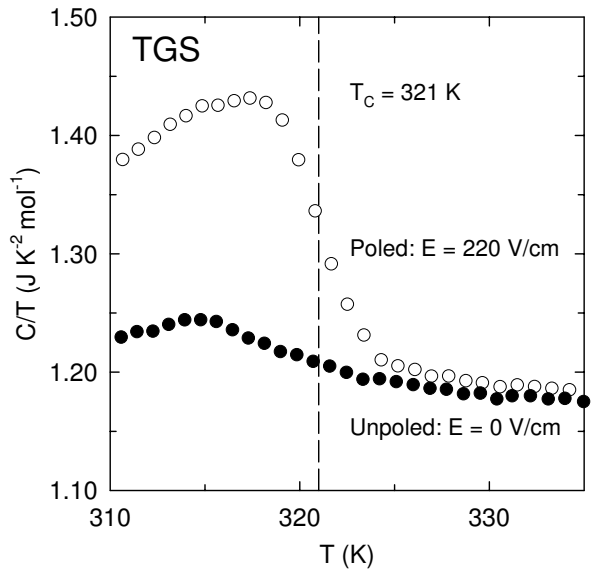}
    \caption{Temperature dependence of $C$ near $T_C$ as a
    function of electric field, 0~V/cm (solid circles) and
    220~V/cm (open circles).}
    \label{Fig4}
\end{figure}

This result, combined with Lawless's observation~\cite{Lawless2}
that surface excitations significantly contribute to the bulk heat
capacity, suggested that the surface charge emanating from the
pyroelectric coefficient, $dP/dT$, was behaving as though being
shorted by the external magnetic field. In order to test this
hypothesis, TGS crystals were short-circuited by coating all sides
with a thin layer of gold. The same measurements shown in Fig.~1
were made on these gold-coated crystals. Figure~2 shows the
temperature dependence of $C/T$ in the vicinity of $T_C$ for
magnetic fields on short-circuited (gold-plated) TGS crystals in
zero magnetic field (solid squares) and 9~T (open circles) with the
magnetic field applied parallel to the easy {\it b}-axis. The
results show that the $\lambda$ anomaly appears in the absence of a
magnetic field in the short-circuited (gold-plated) TGS crystals.
There is no additional magnetic-field dependence to within
experimental error.

Figure~3 shows the excess specific heat, defined as the quantity
$(C-C_{\rm lat})$, normalized to the universal gas constant, $R$,
versus reduced temperature, $(1-T/T_C)$, on a log-log scale. The
lattice specific heat, $C_{\rm lat}$, was represented by a linear
fit from the ferroelectric phase to the paraelectric phase, as
indicated in the 9~T curve of Fig.~1. The curves above and below
$T_C$ approach the same horizontal line with $n$~=~0 and
~$A$~=~8.1~J K$^{-1}$ mol$^{-1}$ in the expression $(C-C_{\rm
lat})/R=A|1-T/T_C|^n$. As an independent check of the critical
exponents, we note that $RA/T_C$~=~0.21J K$^{-2}$ mol$^{-1}$
(obtained from the fit), is equal to the value of the specific-heat
discontinuity at $T_C$ given by $\Delta C (T_C)/T_C$, shown in Fig.
1, and an exponent of $n$~=~0 signifies a logarithmic singularity
above and below $T_C$~\cite{note2}. The latter result appears to be
consistent with renormalization group theory in which a single
unstable fixed point determines the exponents~\cite{Fisher}.

Lastly, we provide another verification of this effect in electric
field~\cite{note}, as shown in Fig.~4. Measurements of non-grounded
(free-standing) TGS crystals in zero-electric field (solid circles)
and in an applied electric field of 220~V/cm (open circles) are
shown in this figure. Qualitatively mirroring the magnetic field
results shown in Fig.~1, the anomaly is broad near $T_C$ in
zero-electric field, while the characteristic $\lambda$-shape
appears, albeit more broadened than those observed in magnetic
fields, in a modest electric field. In context of our data the
magnetic field dependence appears to be extrinsic in origin. Because
the specific-heat data in magnetic field enabled a determination of
the critical exponents, it is not exactly an artifact. We note that
similar observations involving the magnetic response of space
charges at the surface of a dielectric have been made by
Catalan~\cite{Catalan1}. It was shown that the space charge at
surfaces~\cite{Jim} contributes to an extrinsic magnetic field
effect in dielectrics.

In summary, we have demonstrated that the surface charge on
non-electrically grounded TGS crystals interacts with external
magnetic and electric fields. This effect may be understood on the
basis that the surface charge originating from $dP/dT$ behaves as if
shorted by the external fields. Because many heat capacity
measurements involve measuring a temperature-time derivative, added
contributions originating from $dP/dT$ are easily realized. Using
data obtained in magnetic fields, critical exponents were derived
corresponding to the expression $(C-C_{\rm lat})/R=A|1-T/T_C|^n$,
thereby completing the set for TGS~\cite{Gonzalo}. An exponent of
$n$~=~0 signifies a logarithmic singularity above and below $T_C$.
The ability to effectively "short" the surface charge with an
external magnetic or electric field on free-standing crystals
overcomes experimental difficulties from additional background
subtractions and heat leaks from the electrical grounding leads. Our
observations appear to be consistent with a recent detailed model
that explains the magnetic response of space charge at a surface of
a dielectric material~\cite{Catalan1}.

\begin{acknowledgements}
This work was performed under the auspices of the United States
Department of Energy. We are grateful to Jim Scott, Boris Strukov,
Peter Littlewood, Quanxi Jia, Art Ramirez, and Peter Riseborough.
\end{acknowledgements}


\begin{thebibliography}{100}
\bibitem{note2}
It should be noted that an exponent of $n$~=~0 is also consistent
with a classical set of the critical exponents.
\bibitem{note}
A special platform was constructed to make specific-heat
measurements in an electric field. A puck consisting of gold-plated
glass capacitor plates is placed adjacent to a thermal-relaxation
platform. A dc bias voltage (up to 1 kV) was applied with a power
supply and the platform operates in a Quantum Design PPMS.
\end{thebibliography}
\end{document}